\documentclass{article}
\usepackage{arxiv}
\usepackage{graphicx}
\usepackage{mathtools}
\usepackage{breqn}
\usepackage{float}
\usepackage{latexsym, amsmath, verbatim, graphics,  psfont, epsfig, amssymb, color}
\usepackage{cite}
\usepackage{bm}
\usepackage[caption=false,font=normalsize,labelfont=sf,textfont=sf]{subfig}
\usepackage{textcomp}
\usepackage{stfloats}
\usepackage{url}
\usepackage{subcaption}
\usepackage{csquotes}
\usepackage{booktabs,ragged2e}
\usepackage{bigints}
\usepackage[dvipsnames]{xcolor}
\usepackage{enumitem}
\usepackage{hyperref} 
\usepackage{cleveref} 
\usepackage{pifont}
\usepackage{multicol}
\usepackage{multirow}
\usepackage{threeparttable} 
\def\BibTeX{{\rm B\kern-.05em{\sc i\kern-.025em b}\kern-.08em
    T\kern-.1667em\lower.7ex\hbox{E}\kern-.125emX}}
\usepackage{balance}

\newtheorem{theorem}{Theorem}[section]

\newtheorem{assumption}[theorem]{Assumption}
\newtheorem{proposition}[theorem]{Proposition}
\newtheorem{definition}[theorem]{Definition}
\newtheorem{remark}[theorem]{Remark}

\newtheorem{property}[theorem]{Property}

\allowdisplaybreaks

\let\boldsymbol\bm

\newcommand{\R}{\mathbb{R}}
\newcommand{\Rzero}{\mathbb{R}_{\geq0}}
\newcommand{\N}{\mathbb{N}}
\newcommand{\Nzero}{\mathbb{N}_{\geq 0}}

\newcommand{\K}{\mathcal{K}}

\definecolor{codegreen}{rgb}{0,0.6,0}
\newcommand{\q}{\boldsymbol{q}}
\newcommand{\qd}{\dot{\boldsymbol{q}}}
\newcommand{\qdd}{\ddot{\boldsymbol{q}}}

\newcommand{\norm}[1]{\lVert #1 \rVert}

\newcommand{\bs}[1]{\boldsymbol{ #1 }}
\newcommand{\ol}[1]{\overline{ #1 }}
\newcommand{\ul}[1]{\underline{ #1 }}
\newcommand{\paranthesis}[1]{\left( #1 \right)}

\newcommand{\Max}{\text{max}}

\newcommand{\policy}{\eqref{eq:petc_mechanism}}

\begin{document}

\title{Periodic Event-Triggered Prescribed Time Control of Euler–Lagrange Systems under State and Input Constraints}
\author{\href{https://orcid.org/0000-0002-4270-9443}{\includegraphics[scale=0.06]{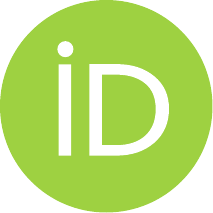}\hspace{1mm}Chidre Shravista Kashyap}  \\ {Center for Cyber-Physical Systems} \\ {IISc, Bengaluru, India} \\ {chidres@iisc.ac.in} \And 
\href{https://orcid.org/0000-0002-9792-1783}{\includegraphics[scale=0.06]{orcid.pdf}\hspace{1mm} Karnan A}\\ Interdisciplinary Data Research Division\\ CSIR-4PI, Bangalore, India. \\ karnan.4pi@csir.res.in 
\And \href{https://orcid.org/0000-0002-5452-8850}{\includegraphics[scale=0.06]{orcid.pdf}\hspace{1mm}Pushpak Jagtap} \\{Center for Cyber-Physical Systems} \\ {IISc, Bengaluru, India} \\ {pushpak@iisc.ac.in}  \And \href{https://orcid.org/0000-0001-8770-2301}{\includegraphics[scale=0.06]{orcid.pdf}\hspace{1mm}Jishnu Keshavan} \\{Dept. of Mechanical Engineering} \\ {IISc, Bengaluru, India} \\ {kjishnu@iisc.ac.in}}
\date{}
	\maketitle
\begin{abstract}
This article proposes a periodic event-triggered adaptive barrier control policy for the trajectory tracking problem of perturbed Euler-Lagrangian systems with state, input, and temporal (SIT) constraints. In particular, an approximation-free adaptive-barrier control architecture is designed to ensure prescribed-time convergence of the tracking error to a prescribed bound while rejecting exogenous disturbances. In contrast to existing approaches that necessitate continuous real-time control action, the proposed controller generates event-based updates through periodic evaluation of the triggering condition. Additionally,  we derive an upper bound on the monitoring period by analysing the performance degradation of the filtered tracking error to facilitate periodic evaluation of the event-triggered strategy. To this end, a time-varying threshold function is considered in the triggering mechanism to reduce the number of triggers during the transient phase of system behaviour. Notably, the proposed design avoids Zeno behaviour and precludes the need for continuous monitoring of the triggering condition. A simulation and experimental study is undertaken to demonstrate the efficacy of the proposed control scheme.
 \end{abstract}

 

\section{Introduction}
Tracking control of Euler-Lagrange (EL) systems has received significant attention with direct relevance to numerous industrial applications \cite{obuz2024robust, he2023ude, lu2019adaptive, he2023output}. Importantly, controller synthesis under stringent SIT constraints plays a vital role in various robotic applications. Accordingly, finite time \cite{feng2023event} or fixed time control \cite{chen2025fixed} techniques have been developed where tracking error convergence is achieved in a finite/fixed time \cite{polykao_FT_EXPAP}. However, in both these cases, a user-defined convergence time cannot be specified such that the tracking error converges in the prescribed settling time. In this regard, prescribed time control (PTC) methods \cite{song2023prescribed} have been employed, where the user can specify the exact convergence time, in contrast with finite and fixed time control studies. 

Most existing results rely on time-triggered control or sampled-data policies, where the controller transmits signals at every sampling interval, regardless of system performance, leading to redundant use of communication and computational resources \cite{zhang2023sampled}. Event-triggered control (ETC) schemes address this issue by updating transmitted signals based on event scheduling \cite{zhang2023recent}, thereby reducing communication and computation demands. Despite the benefits of ETC compared to sampled-data policies, ETC schemes are evaluated at every sampling instant, called continuous event-triggered control (CETC) \cite{zhang2023recent}, which must ensure a positive minimum inter-event time (MIET) to avoid Zeno behaviour \cite{borgers2014event}. Although many theoretical frameworks mitigate this issue, CETC implementation on digital platforms is impractical without advanced hardware and sensors \cite{benitez2020periodic}. To overcome this, discrete event-triggered control based on discrete-time models was proposed in \cite{eqtami2010event}.

Nevertheless, exact discrete-time models for nonlinear systems are often infeasible, and approximations often degrade closed-loop performance \cite{borgers2014event}. To address this limitation, a periodic event-triggered control (PETC) \cite{benitez2020periodic} evaluates the triggering condition at intervals defined by the monitoring period, offering a cost-effective alternative to CETC. However, limited studies ensure prescribed-time convergence of EL systems under PETC while accounting for state and input constraints \cite{song2023prescribed}. Therefore, the proposed study develops an \emph{approximation-free} prescribed-time tracking policy for EL systems, accommodating operating constraints while reducing control communication bandwidth.

\begin{table}[t!] 
	\centering
	\caption{Comparative analysis of the proposed approach with existing studies.}
    \setlength{\tabcolsep}{4pt}
		\begin{tabular}{lccccccc}
			\toprule
			\multirow{2}{*}{Method} & \multirow{2}{*}{ETC} &  \multirow{2}{*}{PETC} & {Approx.} & \multirow{2}{*}{PTC} & State & Input \\
& & & free & & Constraints & Constraints \\ 
\midrule
\cite{benitez2020periodic} & \ding{51}  & \ding{51} & \ding{55} & \ding{55} &  \ding{55} & \ding{55}\\ 
\cite{gao2022novel, dang2024event, kumari2018event,  chen2024robust, chen2024event,peng2023event, gao2022observer}& \ding{51}  & \ding{55} & \ding{55} & \ding{55} &  \ding{55} & \ding{55} \\
\cite{soni2022sliding, zhang2023event} & \ding{51}  & \ding{55} & \ding{55} & \ding{55} &  \ding{51} & \ding{55}\\
\cite{feng2024event}& \ding{51}  & \ding{55} & \ding{55} & \ding{55} &  \ding{51} & \ding{51} \\
\cite{li2022finite} & \ding{51}  & \ding{55} & \ding{51} & \ding{55} &  partial & \ding{55} \\
\cite{liu2021event, sui:2025:FTC} & \ding{51}  & \ding{55} & \ding{51} & \ding{55} &  \ding{55} & \ding{55} \\
\cite{liu2024adaptive} & \ding{51}  & \ding{55} & \ding{51} & \ding{55} &  \ding{51} & \ding{55}\\
\cite{ning:2023:scalar_input, yang:2024:scalar_input, jiang2024event} & \ding{51} & \ding{55} & \ding{55} & \ding{51} & \ding{55} & \ding{55}\\
\cite{chen2023adaptive} & \ding{51}  & \ding{51} & \ding{55} & \ding{51} &  \ding{51} & \ding{55}\\
\cite{Dingxin:2021:icra} & \ding{51} & \ding{55} & \ding{51}  & \ding{51} & \ding{55}& \ding{55} \\
\cite{wang2022adaptive,hu2022event} & \ding{51}  & \ding{55} & \ding{51} & \ding{51} &  \ding{51} & \ding{55}\\
Proposed & \ding{51} & \ding{51} & \ding{51} & \ding{51} & \ding{51} & \ding{51} \\
		\bottomrule
		\end{tabular}%
	\label{table:petc:qualitative_analysis}
\end{table}
The contributions of the proposed methodology are qualitatively compared with leading studies in the design of PETC for EL systems, as shown in Table \ref{table:petc:qualitative_analysis}. The studies in \cite{benitez2020periodic, gao2022novel, dang2024event, kumari2018event,  chen2024robust, chen2024event,peng2023event, gao2022observer, soni2022sliding, zhang2023event, feng2024event} designed ETC for EL systems that depend on model parameters, which are often uncertain or unknown in practice. Alternatively, the studies in \cite{li2022finite, liu2021event, sui:2025:FTC} employ adaptive parametric laws to estimate lumped uncertainties, followed by controller synthesis to achieve better tracking performance. Nevertheless, these studies \cite{benitez2020periodic, gao2022novel, dang2024event, kumari2018event,  chen2024robust, chen2024event,peng2023event, gao2022observer, soni2022sliding, zhang2023event, feng2024event, li2022finite, liu2021event, sui:2025:FTC, liu2024adaptive} do not address the problem of prescribed time convergence for the EL system. To circumvent this limitation, the studies \cite{ning:2023:scalar_input, yang:2024:scalar_input, jiang2024event} designed PTC with nominal model parameters and time-varying gains to ensure prescribed time convergence. In contrast, PTC synthesis with parameter approximation techniques using radial basis functions and time-delayed estimation is studied in \cite{chen2023adaptive} and \cite{Dingxin:2021:icra}, respectively. However, the embedding of state constraints in PTC design is not met, unlike studies \cite{wang2022adaptive,hu2022event} that leverage prescribed performance functions for guaranteeing state constraint satisfaction. Nonetheless, the studies \cite{ning:2023:scalar_input, yang:2024:scalar_input, jiang2024event, chen2023adaptive, Dingxin:2021:icra, wang2022adaptive, hu2022event} do not address the problem of event-triggered control of EL systems under SIT constraints.

Further, concerning MIET, most existing ETC policies \cite{wang2022adaptive, jiang2024event, gao2022novel, zhang2023event} primarily focus on preventing Zeno behaviour but fail to determine MIET explicitly. An additional concern pertains to the judicious selection of a suitable monitoring period based on MIET that enables PETC to capture all events the same way as those in CETC; otherwise, it risks missing events triggered by CETC. Furthermore, static-triggering strategies \cite{liu2024adaptive, chen2023adaptive, wang2022adaptive} impose constant thresholds on the event-triggering mechanism that offer limited flexibility in reducing communication frequency. Alternatively, self-triggering designs proposed in \cite{chen2024robust, gao2022observer} attempt to overcome this by pre-computing event times using MIET upper bounds; however, they execute control actions at fixed instants without verifying triggering conditions, resulting in suboptimal efficiency.

To address these limitations, this study proposes a periodic event-triggered adaptive barrier control policy to ensure robust prescribed-time prescribed-bound convergence of tracking error in EL systems under state and input constraints. Firstly, we leverage the time-based generator (TBG) function that prescribes settling time to construct the filtered tracking error. Then, time-varying inequalities on filtered tracking error are utilised to synthesise a state constraint law to avoid state constraint violations and impose input constraints via a saturation function. Additionally, the triggering mechanism employs a time-varying threshold to reduce communication frequency, and conditions on the monitoring period are derived to limit evaluation of the triggering condition, while preserving the desired tracking error performance. Numerical and experimental validation studies are undertaken to demonstrate the efficacy and superior performance of the proposed scheme compared to leading alternative designs. Thus, the major contributions of this article can be summarized below: 
\begin{enumerate}
	\item An approximation-free periodic event-based adaptive barrier control policy is proposed to reduce communication frequency for tracking control of EL systems under SIT constraints.
    \item In contrast with the studies in \cite{benitez2020periodic, chen2023adaptive}, which fail to address monitoring period selection, this study provides a detailed analysis of its impact within the minimum and maximum inter-event time (IET) bounds. In particular,  by appropriately choosing the monitoring period under the proposed scheme, robust local PTPB convergence of the tracking error is ensured with reduced control updates.
\end{enumerate}
The rest of this article is organised as follows: Section \ref{sec:methodology} details the design of the proposed PETC, then Section \ref{sec:stability_analysis} provides stability analysis, followed by deriving MIET. Then, section \ref{sec:results} discusses the simulation and practical experiment showing the efficacy of the proposed scheme. Finally, the section \ref{sec:conclusion} concludes the discussion of our work.
	
\subsubsection*{Notation} We denote Matrices and vectors by bold letters. Throughout this paper, $\mathbb{R}^n$ and $\mathbb{R}^{n\times n}$ represent the set of all $n-$dimensional real vectors and $n\times n$ real matrices, respectively. The set of all positive (non-negative) integers and reals is denoted by $\N^+$ ($\Nzero$) and $\R^+$ ($\Rzero$), respectively. $\mathbb{S}_+^n~(\mathbb{D}_+^n)$ denotes the set of all $n\times n$ positive definite (positive diagonal) matrices. For a matrix $X$, the notation $X>0$ ($X<0$) symbolizes the positive (negative) definiteness of that matrix. $\bm{I}_n$ denotes the identity of order $n$ and $n$-dimensional  vector having all zeros and ones are denoted by $\bm{0}_n$ and $\bm{1}_n$, respectively. $\norm{\bs{a}}$ denotes the Euclidean norm of a vector $\bs{a}\in\R^n$ and $\norm{\bs{A}}$ represents induced-2 norm for a matrix $\bs{A}\in\R^{m\times n}$. $\text{diag}(\ldots)$ refers for the diagonal matrix. Denote $\N_n=\{1,2,\ldots,n\}$. For vectors $\bm{a}, \bm{b}\in\R^n$, inequality $\bm{a} \preceq \bm{b}$ indicates that $a_i \leq b_i $, for all $i \in \N_n$. For the scalar $x$, the symbols $\ul{x}$ and $\ol{x}$, respectively, denote the lower and upper bounds of the corresponding term.

\section{Methodology} \label{sec:methodology}
\subsection{Problem Statement}
	\noindent The following second-order ordinary differential equation represents the Euler-Lagrange system:
	\begin{align}
		\bm M(\bm q) \qdd+\bm C(\bm q,\bm{\dot q})\bm{\dot q}+\bm G(\bm q)+\bm F(\bm{\dot q})+\bm d(t)=\bm u (t),
		\label{eq:ptcuel_eom}
	\end{align}
which is assumed to satisfy state and input constraints in \eqref {eq:petc_SI_constraints}.
	\begin{align}
	\ul{\bm{q}}	\preceq \bm{q}\preceq \ol{\bm{q}},\qquad 	\ul{\bm{\nu}}	\preceq \dot{\bm{q}}\preceq \ol{\bm{\nu}},\qquad\ul{\bm{u}}\preceq \bm{u}\preceq \ol{\bm{u}},
	\label{eq:petc_SI_constraints}
	\end{align}
where $\q:\R_{\geq0}\rightarrow\R^n$ is the position coordinate, $\qd(t), \qdd(t)$ are the first and second order time differentiation of position $\q(t)$, respectively, 
$\bm{M}(\bm{q}) \in \mathbb{S}_+^n$ is the inertia matrix,
$\bm{C}(\bm{q}, \dot{\bm{q}}) \in \mathbb{R}^{n \times n}$ is the centripetal and Coriolis matrix, $\bm{F}(\bm{q}) \in \mathbb{R}^n$   is damping and friction forces, $\bm{G}(\bm{q}) \in \mathbb{R}^n$ stands for the gravity vector, $\bm{d}(t) \in \mathbb{R}^n$ denotes the external disturbance vector, and $\bm{u}(t) \in \mathbb{R}^n$ is the torque input vector. Additionally, $\ul{\bm{u}}, \ol{\bm{u}}, \ul{\bm{q}},\ol{\bm{q}}, \ul{\bm{\nu}}$, and $\ol{\bm{\nu}} \in\R^n$ are known constant vectors with $i$-th coordinate $\ul{u}_i, \ol{u}_i, \ul{q}_i, \ol{q}_i,\ul{\nu}_i$, and $\ol{\nu}_i$, respectively. For conciseness, arguments of a function are omitted when dependence is evident; e.g., $\bs{C}(\q,\qd)$ and $\bs{d}(t)$ become $\bs{C}$ and $\bs{d}$. The following properties of the EL system \eqref{eq:ptcuel_eom} are invoked in this study:
	\begin{property} \label{prope1}
		The matrix $\dot{\bm{M}}(\bm{q})-2\bm{C}(\bm{q},\dot{\bm{q}})$ is skew-symmetric.
	\end{property}
	\begin{property} \label{prope2}
	There exist $\ul{m},\ol{m}, \ul{l}, \ol{l}\in\mathbb{R}^+$ such that $\ul{m}\bm{I}_n\leq \bm{M}(\bm{q})\leq \ol{m}\bm{I}_n$ and $\ul{l}\bm{I}_n\leq \bm{M}^{-1}(\bm{q})\leq \ol{l}\bm{I}_n$.
	\end{property}
	\begin{property} \label{prope3}
		There exist $\ol{c},\ol{g},\ol{f}\in\mathbb{R}^+$ such that $\norm{\bm{C}(\bm{q},\dot{\bm{q}})}\leq\ol{c}\norm{\dot{\bm{q}}}, \norm{\bm{G}(\bm{q})}\leq\ol{g}$, and $\norm{\bm{F}(\dot{\bm{q}})}\leq\ol{f}\norm{\dot{\bm{q}}}$.  
	\end{property}
The following assumptions about the system dynamics will be exploited in subsequent analysis.
\begin{assumption} \label{ass:parameter} Knowledge of the dynamical terms $\bm{M}, \bm{C}, \bm{F}$, and $\bm{G}$ are not known prior. 
\end{assumption}
\begin{assumption}\label{ass:disturbance}
	There exists a constant $\ol{d}\in\mathbb{R}^+$ such that $\norm{\bm{d}(t)}\leq \ol{d}~\forall t\in\Rzero$.  
\end{assumption}
\begin{assumption} \label{ass:trajectory} For given desired trajectory $\q_r(t),\qd_r(t)$, and $\qdd_r(t)$, there exists constants $\ol{q}_r,\ol{q}_{r_1}, \ol{q}_{r_2}>0$,  such that $\norm{\bm{q}_r}\leq \ol{q}_r,\norm{\dot{\bm{q}}_r}\leq \ol{q}_{r_1}$, and $\norm{\ddot{\bm{q}}_r}\leq \ol{q}_{r_2}$.  Additionally, the trajectory satisfies  $\ul{\bm{q}}	\preceq \bm{q}_r(t)\preceq \ol{\bm{q}}$ and $\ul{\bm{\nu}}	\preceq \dot{\bm{q}}_r(t)\preceq \ol{\bm{\nu}},$  $\forall t\in\Rzero$.
\end{assumption}

For any $\bs{p}\in\R^p$ and radius $r$, define the ball of radius $r$ centered  at $\bs{p}$ as
\begin{equation}
    {\mathcal{B}_r(\bs{p})} = \{\bs{x} \in \R^p : \norm{\bs{x} - \bs{p}} \leq r\},
    \label{eq:ptcuel_prescribed_bound_set}
\end{equation}
and let $|\bs{a}|_{\mathcal{C}}=\inf_{\bs{b}\in{\mathcal{C}}}\norm{\bs{a}-\bs{b}}$ denote the distance of a point $\bs{a}$ from the set $\mathcal{C}\subset\R^p$. The following definition of local prescribed-time prescribed-bound (PTPB) stability allows for a priori selection of the convergence time $T$ and bound $\epsilon$.
\begin{definition}[\cite{csk:2025}]\label{def:ptcuel_petc_PTPB} 
For the system $\dot{\bs{x}} = \bs{f}(t, \bs{x}, \bs{u}, \bs{w})$ with $\bs{x}\in\mathcal{X}$, $\bs{u}=\bs{\mathfrak{u}}(t,\bs{x})\in\mathcal{U}$, and $\bs{w}\in\mathcal{W}$, the equilibrium point $\bs{x}_r\in\mathcal{X}$ is  said to be \emph{local prescribed-time prescribed-bound} (PTPB) stable for user-defined constants $T,\epsilon > 0$, if the closed-loop trajectory $\bs{\Psi}_{\bs{\mathfrak{u}}}(t, t_0, \bs{x}(t_0))$ remains in the ball $\mathcal{B}_\epsilon(\bs{x}_r)$ for any $\bs{x}(t_0)\in\mathcal{C}\subset\mathcal{X}$ after the prescribed time $T$, i.e. $|\bs{\Psi}_{\bs{\mathfrak{u}}}(t, t_0, \bs{x}(t_0))|_{\mathcal{B}_\epsilon(\bs{x}_r)} = 0$ $\forall\ t\geq t_0+T$, where $\mathcal{C}$ is some neighbourhood of $\bs{x}_r$.
\end{definition}
Note that the presence of SIT constraints restricts the viable set of initial conditions as discussed in the study \cite{kgarg_clf}. The study in \cite{csk:2025} further undertakes a detailed feasibility analysis to provide a conservative estimate of this viable set $\mathcal{C}$.

\emph{Problem}: The control objective is to devise an approximation-free periodic event-based adaptive control policy $\bm{u}(t)$ that guarantees robust local prescribed time prescribed bound (PTPB) stable (as defined in \emph{Definition} \ref{def:ptcuel_petc_PTPB}) of the trajectories of the EL system \eqref{eq:ptcuel_eom} with \emph{Properties} \ref{prope1} -- \ref{prope3} and \emph{Assumptions} \ref{ass:parameter} -- \ref{ass:trajectory} under state and input constraints \eqref{eq:petc_SI_constraints}. A further goal is to construct a PETC scheme to reduce the need for constant monitoring of the event-triggering mechanism.

\subsection{Adaptive PETC Design}
In this subsection, we construct a periodic event-triggering mechanism that drives an adaptive barrier function control policy to achieve robust PTPB convergence under state and input constraints.

\begin{figure}
    \centering
    \includegraphics[width=0.6\linewidth]{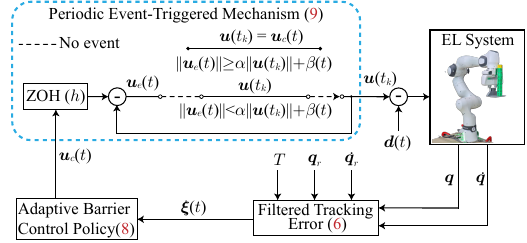}
    \caption{PETC Adaptive Barrier Control Architecture}
    \label{fig:petc_ctrl_arch}
\end{figure}
Firstly, to guarantee the prescribed time stability of the tracking errors, we consider the following time-dependent polynomials $p_1(t)$ and $p_2(t)$ known as TBG \cite{PTC_TBG} to pre-specify settling time $T$: 
\begin{equation}
	\begin{aligned}
	p_1(t)&=
	\begin{cases} -\frac{6}{T^5}t^5+\frac{15}{T^4}t^4 
    -\frac{10}{T^3}t^3+1, &  0\leq t\leq T\\
0, & \text{otherwise}		
	\end{cases},\\
p_2(t)&=
\begin{cases} -\frac{3}{T^4}t^5+\frac{8}{T^3}t^4
-\frac{6}{T^2}t^3+t, & 0\leq t\leq T\\
	0, & \text{otherwise}		
\end{cases}.
\end{aligned} \label{eq:petc_TBG}
\end{equation}
Then tracking error $(\bs{\varepsilon}(t))$ is defined in \eqref{eq:petc_transformed_error} using  \eqref{eq:petc_TBG}.
\begin{flalign}
    \bm{\varepsilon} (t){=} \bm{e} (t){-} (p_1(t)\bm{e}(0) {+} p_2(t)\dot{\bm{e}}(0)),
    \label{eq:petc_transformed_error}
\end{flalign}
with $\bm{e}(t){=} \bm{q}(t){-}\bm{q}_r(t)$, $ \dot{\bm{e}}(t){=}\dot{\bm{q}}(t){-}\dot{\bm{q}}_r(t)$. 

To this end, we have considered the filtered tracking error:
\begin{align}
\bm{\xi}(t)=\dot{\bm{\varepsilon}}(t)+\bm{\K}\bm{\varepsilon}(t),
\label{eq:petc_filter}
\end{align}
where $\bm{\K}\in\mathbb{D}_+^n$ is the gain matrix with $\ul{\K}=\lambda_{\min}(\bm{\K})$ and $\ol{\K}=\lambda_{\max}(\bm{\K})$.

To incorporate the input constraints \eqref{eq:petc_SI_constraints} in the control input $\bm{u}_c$, an unconstrained input variable, $\bm \tau_c\in \mathbb{R}^n$, is introduced with the following relation: $ \bm u_c =\bm{\mathcal{S}}(\bm \tau_c) \bm \tau_c $,
where $\bm{\mathcal{S}}(\bm \tau_c) = \text{diag} ( \mathcal{S}_1(\bm \tau_{c_1}), {\cdot}s, \mathcal{S}_n(\bm \tau_{c_n}) ) \in \R^{n \times n}$  with diagonal entries given by: 
	\begin{flalign}
		\mathcal{S}_i(\tau_{c_i}) = 
		\begin{cases} 
			\ol{u}_i/\tau_{c_i}, & \tau_{c_i} > \ol{u}_i \\
			1, & \ul{u}_i \leq \tau_{c_i} \leq \ol{u}_i \\
			\ul{u}_i/\tau_{c_i}, & \tau_{c_i} < \ul{u}_i
		\end{cases} .\label{eq:petc_saturation_function}
	\end{flalign}
	
An adaptive controller based on a barrier function that accounts for state constraints is taken as follows:
\begin{equation}
    \begin{aligned}
        \bm{\tau}_c &= -\Pi\Upsilon \lceil \bm{\xi} \rfloor^0 - \Pi \|\bm{\chi}\|\norm{\bm{J}} \lceil \bm{\xi} \rfloor^0, \\
    \Pi &= \frac{\rho \|\bm{\xi}\|}{ \omega - \|\bm{\xi}\|},~ \dot{\bm \chi} = -\gamma_1 \bm \chi + \gamma_1 \gamma_2 \bm{J}\bm \xi, ~\bm{\chi}(t_0)=\bm{0}_{2n}, \\
        \Upsilon &= 4 \max \{1, \|\bm{\varepsilon}\|, \|\dot{\bm{\varepsilon}}\|, \|\bm{\varepsilon}\| \|\dot{\bm{\varepsilon}}\|\},
    \end{aligned} \label{eq:petc_control_policy}
\end{equation}
where $\lceil\bs{\xi}\rfloor^0=\bs{\xi}/\norm{\bs{\xi}},\, \rho, \omega,  \gamma_1>0,$ and $0<\gamma_2\leq\norm{\bm{J}}^{-1}$ are  gain parameters, $\bm{J} = [-\bm{I}_n, \bm{I}_n]^\top$. \\

To ensure that the motion planning solution respects physical constraints on joint positions and velocities \cite{zhang2008repetitive}, we have from \eqref{eq:petc_SI_constraints}, let $\ul{\bs{x}}(t) = \ul{\q}-\q_r(t), \ol{\bs{x}}(t) =\ol{\q}-\q_r(t), \ul{\bs{y}}(t) = \ul{\bs{\nu}} - \qd_r(t), \ol{\bs{y}}(t) = \ol{\bs{\nu}}-\qd_r(t),\ul{\bm{e}}_1(t) = \max\{\ul{y}_i(t), \eta(\ul{x}_i(t)-e_i(t))\}$ and $\ol{\bm{e}}_1 = \min\{\ol{y}_i(t), \eta(\ol{x}_i(t)-e_i(t))\}$ with $\eta>0$ and satisfy $\eta>\Max_{i\in\N_n}\{(\ol{\nu}_i-\ul{\nu}_i
)/(\ol{q}_i-\ul{q}_i)\}$. 
Then the time-varying inequality on filtered tracking error is obtained as  $\bm{J}\bm{\xi}\preceq [-\ul{\bm{\psi}}_0^\top,\ol{\bm{\psi}}_0^\top]^\top$, where
	$\ul{\bm{\psi}}_0 = \ul{\bm{\varepsilon}}_1+\bm{\K}\bm{\varepsilon}, \quad \ol{\bm{\psi}}_0 = \ol{\bm{\varepsilon}}_1+\bm{\K}\bm{\varepsilon}$, with $\ul{\bm{\varepsilon}}_1 = \ul{\bm{e}}_1-(\dot{p}_1(t)\bm{e}(0) {+} \dot{p}_2(t)\dot{\bm{e}}(0)$ and $\ol{\bm{\varepsilon}}_1=\ol{\bm{e}}_1-(\dot{p}_1(t)\bm{e}(0) {+} \dot{p}_2(t)\dot{\bm{e}}(0))$. This inequality is then converted into an equality by introducing a set of adaptive, positive-definite unknown auxiliary functions $\bm{\Phi}(t) = [\phi_1^2(t), \phi_2^2(t), \ldots, \phi_{2n}^2(t)]$. Finally, the state constraint equation given in \eqref{eq:petc_control_policy} is obtained by introducing the variable $\bm{\chi} = \bs{J}\bs{\xi} + \bs{\Phi} - [-(\ul{\bs{\psi}}_0+\sigma\bs{1}_n)^\top,\  (\ol{\bs{\psi}}_0 - \sigma\bs{1}_n)^\top]^\top$, where $\sigma>0$ is a user-prescribed safety margin. The unknown functions $\phi_i(t), i\in\mathbb{N}_{2n}$ can be obtained similar to that in study \cite{csk:2025}. 

To reduce the transmission frequency of control signals and eliminate the necessity for continuous monitoring, the \emph{PETC} mechanism is defined as
\begin{equation}
    \begin{aligned}
    \bm{u}(t) &= \bm{u}_c(t_k) \quad \forall t \in [t_k, t_{k+1}), \; k \in \mathbb{N}_{\geq 0}, \\
    t_{k+1} &= \min_{l{\in}\mathbb{N}^+}\{lh {>} t_k \mid \norm{\bm{u}_e(lh)} {\geq} \alpha \norm{\bm{u}(t)} {+} \beta(lh)\}, \\
    \beta(t)&=\begin{cases}
				\beta_0(T-t), &  t<T\\
				0, &  t\geq T
			\end{cases},
    \end{aligned} \label{eq:petc_mechanism} 
\end{equation}
where $\bm{u}_e(t)=\bm{u}_c(t) - \bm{u}(t)$, $h$ is the monitoring period, $0<\alpha<1$ and $0<\beta_0<1$ are positive design parameters, time $t$ with subscript $k$ indicates the event instant at $t = t_k$, which will be derived later.  Once the mechanism \eqref{eq:petc_mechanism} is triggered, the control input $\bm{u}$ will be updated by the control law $\bm{u}_c$ at $t_{k+1}$. Thus, for $t \in [t_k, t_{k+1})$, $ \bm{u}(t) = \bm{u}_c(t_k)$, and $\bs{u}(t)$ updated at the moment it violates the following inequality \eqref{eq:petc_not_triggered_value}: 
\begin{equation}
    \norm{\bm{u}_c(t)-\bm{u}(t_k)} < \alpha \norm{\bm{u}(t_k)} + \beta(t). \label{eq:petc_not_triggered_value}
\end{equation}
An illustration of this PETC mechanism is shown in Fig. \ref{fig:petc_ctrl_arch}.
\begin{remark}
The frequency of control policy updates decreases for increasing values of $\alpha$ and $\beta_0$, resulting in improved energy and computational efficiency, but it makes the system less sensitive to smaller disturbances, thus leading to degraded tracking performance. Alternatively, smaller values of $\alpha$ and $\beta_0$ result in more frequent triggering, which makes the system more responsive to disturbance rejection but requires more computation. Furthermore, introducing the time-varying function $\beta(t)$ enhances flexibility in the ETC mechanism \eqref{eq:petc_mechanism} by dynamically adjusting the triggering threshold, unlike the fixed relative threshold strategy in \cite{chen2023adaptive}. In particular, $\beta(t)>0$ for $t<T$ reduces unnecessary triggers during the transient phase, while $\beta(t)=0$ for $t\geq T$ increases trigger frequency near the reference, ensuring responsiveness and accurate convergence.
\end{remark}
\begin{remark} \label{rem:PETC}
Extending CETC to PETC involves verifying the event condition \eqref{eq:petc_mechanism} after a time $t=t_k+h$, with the monitoring period $h$ affecting control update frequency and tracking performance. Thus, designing system-dependent admissible bounds $\nu,\,h^*>0$ satisfying $\nu<h<h^*$ becomes important to limit control update frequency even while guaranteeing robust tracking performance. In particular, for the choice of MIET being $\nu$, setting $h<\nu<h^*$ ensures that the triggering condition \eqref{eq:petc_mechanism} captures all events if continuous evaluation of \eqref{eq:petc_not_triggered_value} at each triggering instant is considered, resulting in smooth tracking performance. In contrast, $h\in(\nu,h^*)$ causes an increase in actuation error $\bs{u}_e$ resulting in tracking performance degradation. In this regard, the computation of $\nu$ and $h^*$ is addressed in \emph{Theorem} \ref{thm:petc_miet} and \emph{Proposition} \ref{thm:petc_upper_bound_h}, respectively.

\end{remark}
	\section{Stability Analysis} \label{sec:stability_analysis}
    This section provides a detailed stability analysis based on the PETC condition \eqref{eq:petc_mechanism} through \emph{Theorem} \ref{thm:petc_stab}, \emph{Theorem} \ref{thm:petc_miet}, and \emph{Proposition} \ref{thm:petc_upper_bound_h}.
	\begin{theorem}
	    Consider the state and input-constrained EL system \eqref{eq:ptcuel_eom} with \emph{Properties} \ref{prope1}-\ref{prope3} and satisfying \emph{Assumptions} \ref{ass:parameter}-\ref{ass:trajectory}. Then the adaptive control policy \eqref{eq:petc_control_policy} under the event-triggered mechanism \eqref{eq:petc_not_triggered_value} ensures that local PTPB stability is achieved within the prescribed time $T$, with the prescribed bound given by $\epsilon = \paranthesis{(\omega/\ul{\K})^2+\omega(\ol{\K}/\ul{\K}+1))^2}^{1/2}$.  \label{thm:petc_stab}
	\end{theorem}

	 \emph{Proof}: 
	To prove the stability of the EL system \eqref{eq:ptcuel_eom} under the control policy \eqref{eq:petc_control_policy}, consider the following
	Lyapunov function:
	\begin{align}
		\mathcal{V}(t) = \frac{1}{2} \bm{\xi}^\top \bm M \bm{\xi} + \frac{1}{2\gamma_1} \bm{\chi}^\top \bm{\chi}.
		\label{eq:V}
	\end{align}
	Then, with ETC policy,  the time derivative of Lyapunov function \eqref{eq:V} along the system \eqref{eq:ptcuel_eom} yields
	\begin{align}
		\dot{\mathcal{V}} 
		&= \bm{\xi}^\top ( \bm{u}+ \bm{\delta}) + \frac{1}{2} \bm{\xi}^\top (\dot{\bm{M}} - 2\bm{C})\bm{\xi} + \frac{1}{\gamma_1} \bm{\chi}^\top \dot{\bm{\chi}}, \label{eq:Vdot}
	\end{align}
	
	where $\bm{\delta}(t)$ is the lumped uncertainty taken as
	\begin{equation}
		\begin{aligned}
		    \bm{\delta}(t) = &\bm{C}(\bm{q}, \dot{\bm{q}})(\bm{\xi} - \dot{\bm{q}}) + \bm{M}(\bm{q})(-\ddot{\bm{q}}_r - \ddot{\bm{e}}_r + \bm{\K}\dot{\bm{e}}) \\
         &- \bm{G}(\bm{q}) - \bm{F}(\dot{\bm{q}}) - \bm{d}(t).
		\end{aligned}
	\end{equation}
	Using \emph{Properties} \ref{prope1}--\ref{prope3}, \emph{Assumption} \ref{ass:disturbance} of EL system \eqref{eq:ptcuel_eom}, and equation \eqref{eq:petc_filter}, an upper bound on $\bm{\delta}(t)$ can be computed  as
		\begin{align}
			\|\bm{\delta}\| 
            \leq \ol{\delta}\Upsilon,  \label{eq:bound on uncer}
		\end{align}
	where $\ol{\delta} = \max\{\delta_0, \delta_1, \delta_2, \delta_3\}$  with	
    \begin{flalign*}
		 \delta_1&=\ol{m}(\ol{q}_{r_2} + \ol{e}_{r_2}) + \ol{c}(\ol{q}_{r_1} + \ol{e}_{r_1})^2 + \ol{f}(\ol{q}_{r_1} + \ol{e}_{r_1}) + \ol{g} + \ol{d}, \\
		\delta_2 &= \ol{c}(\ol{q}_{r_1} + \ol{e}_{r_1})\ol{\K}, 
		\delta_3 = \ol{m}\ol{\K} + \ol{c}(\ol{q}_{r_1} + \ol{e}_{r_1}) + \ol{f},~ \delta_0 = \ol{c}\ol{\K}, \\
        \ol{e}_{r_1} &= \underset{t\geq0}{\Max}\ (\dot{p}_1(t)\bm{e}(0) + \dot{p}_2(t)\dot{\bm{e}}(0)),\\ \ol{e}_{r_2} &= \underset{t\geq0}{\Max}\ (\ddot{p}_1(t)\bm{e}(0) +\dot{p}_2(t)\ddot{\bm{e}}(0)),
    \end{flalign*}

Equation \eqref{eq:Vdot} can be reduced to the following using the control input \eqref{eq:petc_control_policy} and \emph{Property} \ref{prope1} as:
	\begin{align*}
		\dot{\mathcal{V}}
		=& -\frac{\rho \Upsilon}{\omega - \norm{\bm{\xi}} }\bm{\xi}^\top \bm{\mathcal{S}}  \bm{\xi}  - \frac{\rho \norm{\bm{\chi}}\norm{\bm{J}}}{\omega - \norm{\bm{\xi}} }\norm{\bm{\xi}}^\top  \bm{\mathcal{S}} \bm{\xi}\\&+\bm{\xi}^\top\paranthesis{\bm{u}-\bm{u}_c}+ \bm{\xi}^\top \bm{\delta} - \bm{\chi}^\top \bm{\chi} + \gamma_2 \bm{\chi}^\top \bm{J} \bm{\xi}.
	\end{align*}
	Since  $\norm{\bm{u}}\leq \ol{u}$, therefore, using inequality \eqref{eq:petc_not_triggered_value}, we have
	\begin{equation}
		\norm{\bm{u}_c-\bm{u}} < \alpha \norm{\bm{u}} + \beta(t)< \ol{u}+ T=\kappa. 
		\label{eq:bound on u}
	\end{equation}
	Because of the density property of $\mathbb{R}$, there exists $r > 0$ such that $0 < r \leq \lambda_{\min}\{\bm{\mathcal{S}}\} < 1$. Moreover, we have $\Upsilon \geq 4$. Using \eqref{eq:bound on uncer} and \eqref{eq:bound on u},  the above inequality can be written as
	\begin{align*}
		\dot{V} {<}{-}4 \left( \frac{ r \rho \norm{\bm{\xi}}  }{\omega{-}  \norm{\bm{\xi}} } {-} \ol \delta -\frac{ \kappa }{4}\right)   \norm{\bm{\xi}}{-} \left( \frac{ r \rho \norm{\bm{\xi}}  }{\omega{-}  \norm{\bm{\xi}} } {-} \gamma_2 \right)   \norm{\bm{J}}   \norm{\bm{\chi}}  \norm{\bm{\xi}}. 
	\end{align*}
	Then, provided $ \norm{\bm{\xi}} > \text{max} \left\{\frac{\left(\ol{\delta} + \frac{\kappa}{4}\right) \omega}{r \rho +  \left(\ol{\delta} + \frac{\kappa}{4}\right)}, \frac{\gamma_2 \omega}{r \rho + \gamma_2} \right\}$, we can prove that $\dot{\mathcal{V}}<0.$ As $\bm{\xi}(0)=\bm{0}_n$, it follows that
	\begin{align}
		\norm{\bm{\xi}}<\max\left\{\frac{\left(\ol{\delta} + \frac{\kappa}{4}\right) \omega}{r \rho +  \left(\ol{\delta} + \frac{\kappa}{4}\right)}, \frac{\gamma_2 \omega}{r \rho + \gamma_2}\right\} < \omega, \forall t\geq 0. \label{norm of chi}
	\end{align}
Then, using the filtered tracking error in \eqref{eq:petc_filter}, upon solving the first-order ODE, one can obtain 
\begin{align}
	\norm{[\bm{e}^\top, \dot{\bm{e}}^\top]^\top}<((\omega/\ul{\K})^p+(\omega(\ol{\K}/\ul{\K}+1))^p)^{1/p}.
\end{align} 
Moreover, similar to the proof of \cite[Theorem 2]{csk:2025}, adherence to joint angle and velocity constraint limits can be demonstrated for the choice of a user-defined safety margin $\sigma=\omega$. Consequently, the closed-loop trajectories of system \eqref{eq:ptcuel_eom} under control policy \eqref{eq:petc_control_policy} achieve local PTPB stability as in \emph{Definition} \ref{def:ptcuel_petc_PTPB} despite external disturbances. \hfill $\blacksquare$
	
\begin{remark} \label{rm2}
    Note that from \eqref{norm of chi}, one can deduce that $	\norm{\bm{\xi}}< a \omega, $ with $a=\max\left\{(\ol{\delta} + \frac{\kappa}{4})/(r \rho +  (\ol{\delta} + \frac{\kappa}{4})), \gamma_2/(r \rho + \gamma_2)\right\}<1$. Therefore,  the adaptive gain $\Pi$ in \eqref{eq:petc_control_policy} has an upper bound of $\rho a/(1-a)$. Consequently, the adaptive gain $\Pi(t)$ in \eqref{eq:petc_control_policy} is singularity-free and remains well-defined $\forall t\geq 0$.
\end{remark}	
\subsection{Analysis of the Periodic Event-Triggering Mechanism}
Closed-loop system performance is sensitive to the choice of monitoring period. In particular, the PETC mechanism behaves similarly to the CETC if the monitoring period is less than the MIET, but potentially loses triggering events otherwise. Moreover, exceedingly large monitoring periods cause performance degradation. To address this, an upper bound on the monitoring period is derived, in addition to the computation of a lower bound on MIET to avoid non-Zeno behaviour.

\begin{theorem}[minimum inter-event time] There exists $\nu>0$ such that the inter-event time (IET) generated by the event-triggered mechanism  \eqref{eq:petc_mechanism} are lower bounded by $\nu$ , i.e., $t_{k+1}-t_k>\nu~\forall k\in\mathbb{N}_{\geq 0}$. \label{thm:petc_miet}
		   \end{theorem}

\emph{Proof}: 
Finding derivative of $\norm{\bm{u}_e(t)}$, we get
\begin{align}	\frac{d}{dt} \norm{\bm{u}_e(t)} = \frac{\bm{u}_e(t)^T \left(\dot{\bm{u}}_c(t) - \dot{\bm{u}}(t)\right)}{\norm{\bm{u}_e(t)}} \leq \norm{\dot{\bm{u}}_c(t)}. \label{eq:ue_dot_bound}
\end{align}
Also, from equation \eqref{eq:petc_control_policy}, we have 
\begin{align}
    \dot{\bm u}_c(t) = \dot{\bm{\mathcal{S}}}(\bm \tau_c(t)) \bm \tau_c(t)+\bm{\mathcal{S}}(\bm \tau_c(t)) \dot{\bm \tau}_c(t). \label{eq: ucdot}
\end{align}
where $\dot{\bm{\tau}}_c(t) = -\dot{K}_1(t)\Upsilon -K_1(t)\frac{d}{dt} \Upsilon- \dot{K}_1(t) \|\bm{\chi}\|\norm{\bm{J}}- K_1(t) \frac{d}{dt}\Big(\|\bm{\chi}\|\Big)\|\bm{J}\|$ 
    with $K_1(t)=\frac{\rho \bm{\xi}}{ \bm{\omega} - \|\bm{\xi}\|}$. 
    
Computing the norm of $\dot{K}_1(t)$ (from \emph{Remark} \ref{rm2}, there exists $0<a_0<a$ such that $\norm{\bs{\xi}}=a_0\omega$), we get
\begin{align}
    \norm{\dot{K}_1(t)}\leq\frac{\rho\omega\norm{\dot{\bm{\xi}}}}{(\omega-\norm{\bm{\xi}})^2} < \frac{\rho l_1}{(1-a_0)^2} := l_2. \label{eq:bound of Kdot}
\end{align} 
where $l_1=\ol{l}\,  \ol{u}+\ol{l}\,\ol{c}\,\ol{\nu}\,\omega+\ol{l}\,\ol{\delta}$ can be obtained as
\begin{align}		   		\norm{\dot{\bm{\xi}}}\leq\norm{\bm{M}^{-1}}\paranthesis{\norm{\bm{u}}+\ol{c}\norm{\dot{\bm{q}}}\omega+\norm{\bm{\delta}}}\leq l_1. \label{eq:bound on chidot}
\end{align} 
Also, $ 
    |{\frac{d\norm{\bm{\chi}}}{dt}}| \leq\norm{\dot{\bm{\chi}}}<\gamma_1\omega+\gamma_1\gamma_2\norm{\bm{J}}\omega:=l_3$. 
    
From the solution for the tracking error \eqref{eq:petc_filter}  with TBG  $\bm{\varepsilon}(t)$  we can obtain 
$
    \left| \frac{d}{dt} \Upsilon \right| \leq  4 (\|\dot{\bm{\varepsilon}}(t)\|^2 + \|\bm{\varepsilon}(t)\| \|\ddot{\bm{\varepsilon}}(t)\|)=4l_4$, 
where $l_4{=}({\omega l_1}/{\ul{\K}})+\omega^2({1+{\ol{\K}}/{\ul{\K}}})({({1+{\ol{\K}}/{\ul{\K}}}+{\ol{\K}}/{\ul{\K}})})$. 

Making use of the inequalities derived above, we have the following bound:
\begin{align}
    \frac{d}{dt} \|\bm{u}_e(t)\|< l_M, \label{eq:petc_uedot_bound}
\end{align}
    where $l_M=l_2\paranthesis{4+\sqrt{2}\omega}+\paranthesis{{\rho}/(1-a)}\paranthesis{4l_4+\sqrt{2}l_3}$. 
Taking integral on both sides of the ETC condition \eqref{eq:petc_mechanism} over the interval $[t_k, t_{k+1})$ and using the inequality \eqref{eq:petc_uedot_bound} we have
\begin{align}
    t_{k+1}-t_{k}> \frac{\alpha \norm{\bm{u}(t)}+ \beta(t)}{l_M}>\frac{\alpha \ol{u}+\beta_0 T}{l_M}:=\nu.
    \label{eq:petc_MIET}
\end{align}
Therefore, there exists $\nu>0$ such that $t_{k+1}-t_k>\nu$ resulting in non-Zeno behaviour in the PETC mechanism \eqref{eq:petc_mechanism}. \hfill $\blacksquare$

\begin{proposition} \label{thm:petc_upper_bound_h} 
Let $\omega > 0, a\in[0,1)$, and $l_1 > 0$ such that $\|\dot{\bs{\xi}}(t)\| \leq l_1$. Then, prescribed tracking performance $||\boldsymbol{\xi}(t)||<\omega,\,\forall\, t\geq t_0$ is ensured provided the monitoring period $h$ in the PETC mechanism \eqref{eq:petc_mechanism} satisfies the inequality $h<h^*=(1-a)\omega/l_1$.
\end{proposition}
\noindent\emph{Proof}:
Firstly, from \emph{Remark} \ref{rm2} we have $\norm{\bs{\xi}(t_k)} < a\omega$. Then, to compute the upper bound on the monitoring period $h$, consider the time instant $t$ at which $\norm{\bs{\xi}(t)}=\omega$, so that
\begin{flalign}
    \omega=\norm{\bs{\xi}(t)} &\leq \norm{\bs{\xi}(t_k)} + \int_{t_k}^{t} \norm{\dot{\bs{\xi}}(s)}ds \nonumber \\
    & < \norm{\bs{\xi}(t_k)} + l_1(t-t_k), \ \forall\ t\in[t_k,t_{k+1})\nonumber \\
    \implies t-t_k &> \frac{(1-a)\omega}{l_1}.
    \label{eq:petc_MXIET}
\end{flalign}
Thus, the upper bound on the monitoring period is chosen as $h = t-t_k< h^*$. \hfill$\blacksquare$

\begin{remark}
    From \eqref{eq:petc_MIET} and \eqref{eq:petc_MXIET}, it is evident that the bounds $\nu$ and $h^*$ are derived based on system parameter properties (see \emph{Properties} \ref{prope1} -- \ref{prope3}) and controller parameters $\alpha,\,\beta_0$. Note that choosing a lower value of $\omega$ results in a smaller prescribed bound (given by  \emph{Theorem} \ref{thm:petc_stab}) and a lower value for the upper bound on the monitoring period $h^*$. Thus, better tracking accuracy can only be achieved at the expense of more frequent evaluation and triggering of control signals (even while avoiding Zeno behaviour), thus ensuring robust PTPB tracking performance under SIT constraints.
\end{remark}

\section{Results and Discussion} \label{sec:results}
This section presents simulation results while comparing with leading studies \cite{wang2022adaptive} and \cite{chen2023adaptive} to demonstrate the superior performance of the proposed scheme \policy{} along with the PETC mechanism \eqref{eq:petc_mechanism}. In addition, experimental studies on a 7-DoF serial manipulator are also undertaken to showcase the validity of the proposed scheme.
\subsection{Simulation results}
\begin{figure*}[!t] 
    \centering
    \subfloat [Joint Angular Position $(\q)$]
    {\includegraphics[width=0.325\textwidth]{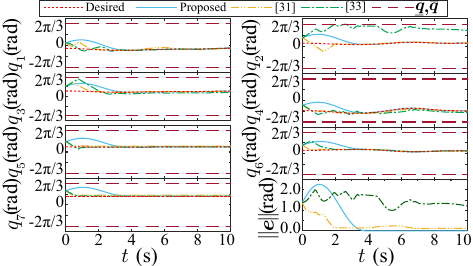}%
    \label{fig:petc_iiwa14_jointPosition}}
    \hfill
    \subfloat [Joint Angular Velocity $(\qd)$]
    {\includegraphics[width=0.325\textwidth]{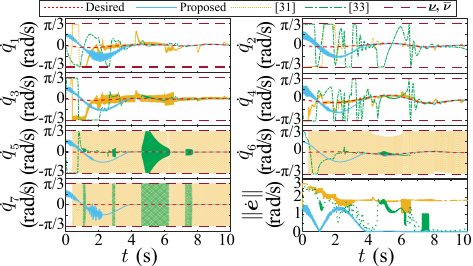}
    \label{fig:petc_iiwa14_jointVelocity}}
    \hfill
    \subfloat[Control Input $(\bs{u})$]
    {\includegraphics[width=0.325\textwidth]{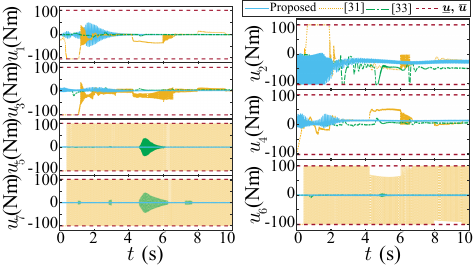}
    \label{fig:petc_iiwa14_jointTorque}}
    \caption{Simulation results for the IIWA 14 robot depicting trajectory tracking under the proposed control policy \policy{} with comparisons against the studies in \cite{chen2023adaptive, wang2022adaptive} for the chosen prescribed time of $T = 4\text{s}$ and $h=0.0002$s.}
\end{figure*}

This subsection illustrates the efficacy of the proposed control scheme \policy{} using a simulation comparison for KUKA LBR IIWA 14 R820 (IIWA 14) with an adaptive event-triggered prescribed-time control scheme \cite{wang2022adaptive}, and adaptive prescribed settling time periodic ETC \cite{chen2023adaptive}  are considered in this study.

For the IIWA 14 robot, the reference trajectories are derived from the tricuspid trajectory in Cartesian space. The simulation step size is $0.1$ms for all numerical experiments considered in the study.
The control parameters for \cite{wang2022adaptive} are chosen as $a {=} 10^{-5}$, $b = 10^{-4}$, $\alpha {=}0.25$, $\epsilon_1 {=} 0.3$, $\epsilon_2{ =} 10^{-3}$, $\sigma_{a_1}{=} \sigma_{a_2}{=} 0.5$, $\sigma_{\theta_1}{=} \sigma_{\theta_2} {=} 3$, $k_c {=} 2\pi/3$, $\Delta k_c {=} 0$, $\gamma_1 {=} \gamma_2 {=} \beta$ and $\beta_1{=}\beta_2 {=} \alpha$. For the \cite{chen2023adaptive} we choose $\mu {=} 0.0029$, $d {=} [0.0241, 0.0241, 0.0241]^\top$, $\beta {=} 0.01$, $\sigma {=} 1$, $\chi {=} 0.2$, $\nu {=} 2.6$, $\rho {=} 0.8$, $c_1 {=} 10$,  $c_2 {=} 26$,  $k_{c_1} {=} 2\pi/3{\cdot}\bm{1}_7$, and $k_{c_2}  {=} \pi/3{\cdot}\bm{1}_7$. For the proposed method $\bm{\mathcal{K}} {=} \text{diag}(1600, 8000, 2200, 4000, 800, 1200, 128)$, $\rho {=} 12.5$, $\omega {=} 25$, $\gamma_1 {=} 1$, $\gamma_2 {=} 0.4$, $T {=} 4$s, $\alpha{=}0.0029$, $\beta_0 {=} 0.0241, h=0.0002$s,  state and input constraints are set as $\ul{\bm{q}} {=} {{-}\ol{\bm{q}}} {=} 2\pi/3{\cdot}\bm{1}_7$, $\ul{\bm{\nu}} {=} {{-} \ol{\bm{\nu}}} {=} \pi/3{\cdot}\bm{1}_7$, $\ul{\bm{u}} {=} {-} \ol{\bm{u}} {=} 100{\cdot}\bm{1}_7$Nm. It is assumed that the external disturbance $d(t) {=} [0.02,0.1,0.1,0.1, 0.02, 0.02, 0.002]^\top \text{Nm}$. 

Fig.  \ref{fig:petc_iiwa14_jointPosition} and  \ref{fig:petc_iiwa14_jointVelocity} show that the proposed control policy \policy{} drives the IIWA 14's joint position and velocity to the reference trajectory within the prescribed settling time $T = 4$s, and Table \ref{tab:petc_iiwa14_comparison_steady_State_metrics} shows that the steady-state performance for the IIWA 14's joint position and velocity is in order of $0.001$ deg and $0.1$ deg/s, respectively, illustrating the efficacy of the proposed policy. However, the control laws in the studies \cite{chen2023adaptive} and \cite{wang2022adaptive} deliver inferior trajectory tracking accuracy in both joint position and velocity compared to the proposed controller \policy{}, as seen in Table \ref{tab:petc_iiwa14_comparison_steady_State_metrics}, which is further evident from the plots of $\norm{\bs{e}(t)}$ and $\norm{\dot{\bs{e}}(t)}$ in Figs. \ref{fig:petc_iiwa14_jointPosition} and \ref{fig:petc_iiwa14_jointVelocity} respectively. The inadequate tracking performance of the controllers in \cite{chen2023adaptive} and \cite{wang2022adaptive} may arise from the comparatively higher chatter of the control signal. A possible reason for these undesirable oscillations of the input signal could be that the studies \cite{chen2023adaptive} and \cite{wang2022adaptive} do not incorporate the SIT constraints in the controller structure (see Table \ref{table:petc:qualitative_analysis}), thus rendering it difficult to deliver robust tracking performance under such constraints. Furthermore, Table \ref{Tab:petc_comparison_petc_metrics} shows that the proposed control policy \policy{} with the PETC mechanism \eqref{eq:petc_mechanism} requires only $6.2\%$ of updates to the control signal to achieve robust tracking performance compared to studies \cite{chen2023adaptive} and \cite{wang2022adaptive} that require $32.88\%$ and $56.05\%$, respectively. Although studies \cite{chen2023adaptive} and \cite{wang2022adaptive} have higher minimum and maximum IET  with $0.0003$s and $0.0524$s, respectively, compared to the proposed PETC adaptive barrier policy, the average release period and transmission percentage show that superior tracking performance is achieved with far fewer control updates compared to these reference designs. Moreover, there is no existence of Zeno-behaviour with MIET = $0.0002$s. Consequently, the proposed methodology performs significantly better in following a reference trajectory while satisfying the SIT constraints.

\begin{table}[h!]
\centering
\caption{Comparison of joint position (in deg) and velocity (in deg/s) RMSE in steady state $(t\geq T=4s)$ between the proposed method, \cite{chen2023adaptive}, and \cite{wang2022adaptive} for the IIWA 14 robot with $T = 4$s and $h=0.0002$s.}
\renewcommand{\arraystretch}{1.2}
\begin{tabular}{|c|c|c|c|c|c|c|c|}
    \hline
     Method & $q_1 $ & $q_2 $ & $q_3 $ & $q_4 $&$q_5 $ & $q_6 $ & $q_7 $ \\

    \hline
       Proposed & \textbf{0.004} & \textbf{0.026} & \textbf{0.007} & \textbf{0.029} & \textbf{0.002} & \textbf{0.002} & \textbf{0.000} \\ \hline
       \cite{chen2023adaptive} & 3.865 & 1.324 & 1.052 & 2.916 & 3.162 & 1.154 & 4.049 \\ \hline
       \cite{wang2022adaptive} & 0.678 & 59.002 & 6.236 & 7.970 & 0.086 & 1.108 & 0.006  \\
    \hline
    Method & $\dot{q}_1 $ & $\dot{q}_2 $ & $\dot{q}_3 $ & $\dot{q}_4 $&$\dot{q}_5 $ & $\dot{q}_6 $ & $\dot{q}_7 $ \\
  
    \hline
    Proposed & \textbf{0.4} & \textbf{0.4} & \textbf{0.2} & \textbf{0.4} & \textbf{0.1} & \textbf{0.8} & \textbf{0.1} \\
\hline
 \cite{chen2023adaptive} & 6.0 & 2.6 & 5.9 & 4.1 & 46.4 & 44.7 & 46.4 \\
 \hline
 \cite{wang2022adaptive} & 2.1 & 19.9 & 2.1 & 11.3 & 11.7 & 1.8 & 26.0 \\
    \hline
\end{tabular}
\label{tab:petc_iiwa14_comparison_steady_State_metrics}
\end{table}

\begin{table}[h!]
	\centering
		\caption{Comparison of IET values between the proposed PETC scheme, \cite{chen2023adaptive} and  \cite{wang2022adaptive}, for the IIWA 14  robot with $T=4$s and $h=0.0002$s.}
\begin{threeparttable}[b]
		\begin{tabular}{|c|c|c|c|c|}
			\hline
			\multirow{2}{*}{Method} & \multirow{2}{*}{Transmission\tnote{a} (\%)} & Average Release & \multicolumn{2}{c|}{IET} \\ \cline{4-5}
         & & Period\tnote{b} (s) & Min. & Max.\\\hline

			Proposed & \textbf{6.20} & \textbf{0.0016} &  0.0002 & {0.0098}  \\\hline
			\cite{chen2023adaptive} & 32.88 & {0.0003} & \textbf{0.0003} & 0.0006 \\\hline
			\cite{wang2022adaptive} & 56.05 & 0.0002 & 0.0001 & \textbf{0.0524} \\\hline
	\end{tabular}
    \begin{tablenotes}
        \item [a] Transmission (\%) = $\frac{\text{Number of events}}{\text{Total number of sample instants}}\times100$.
        \item [b] Average Release Period  $=\paranthesis{\sum\limits_{k=1}^{n_e}t_k}/{n_{e}}$, $n_e=$Total number of events.
    \end{tablenotes}
\end{threeparttable}
	\label{Tab:petc_comparison_petc_metrics}
\end{table}

\subsection{Experimental Studies}
\begin{figure}
    \centering
    \includegraphics[width=0.5\linewidth]{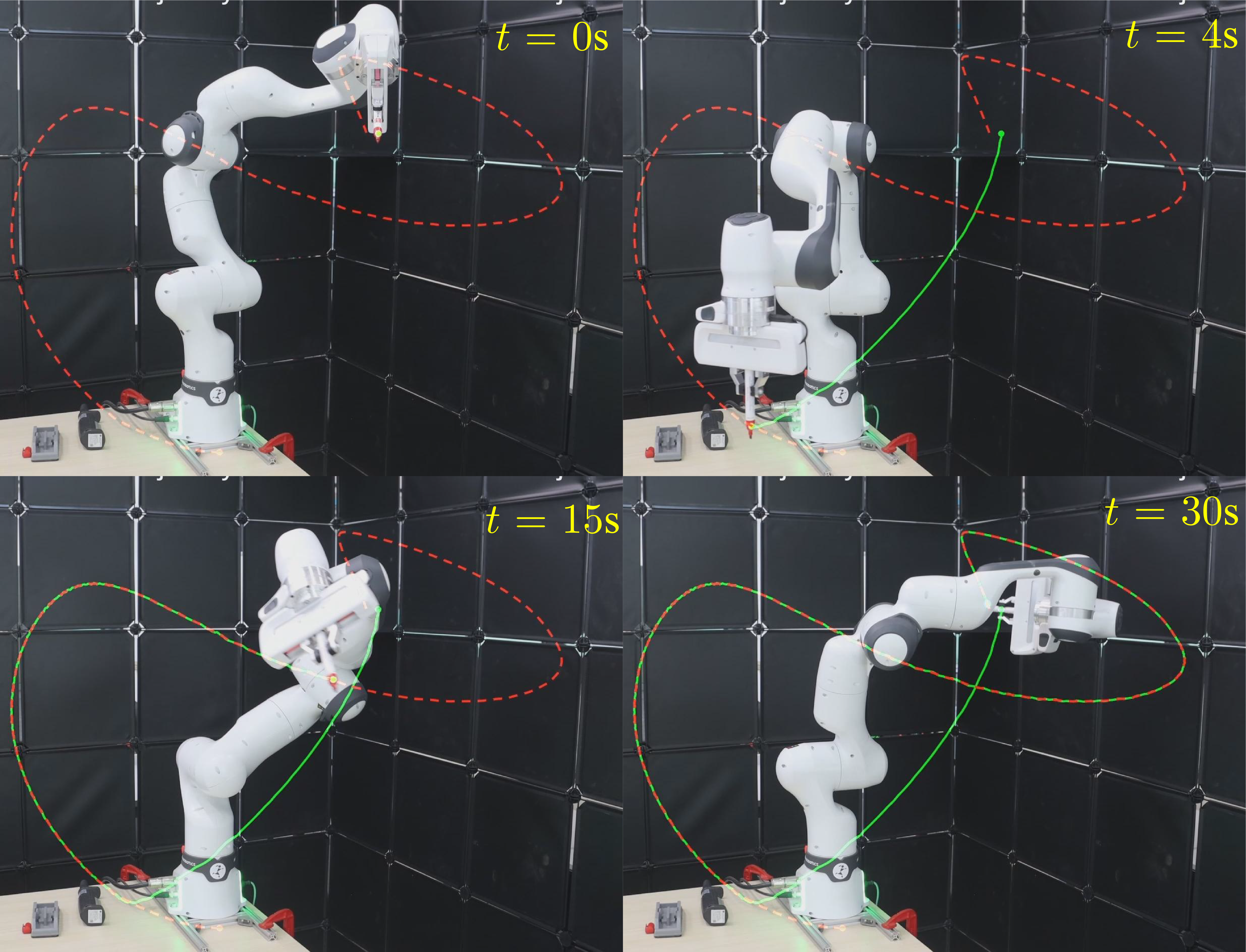}
    \caption{The FR3 robot arm tracking the desired trajectory within the prescribed time $T=4s$. The red dotted and green solid lines denote the desired and traced trajectories respectively.}
    \label{fig:petc_franka_illustration}
\end{figure}
\begin{table}[!ht]
    \centering
    \caption{Experimental evaluation of the proposed policy under PETC mechanism \eqref{eq:petc_mechanism} for the FR3 manipulator arm.}
    \begin{tabular}{|c|c|c|c|c|}
    \hline
       Monitoring & \multirow{2}{*}{Transmission (\%)} & Average Release & \multicolumn{2}{c|}{IET} \\ \cline{4-5}
        Period($h$) & & Period (s) & Min. & Max.\\\hline
        \multicolumn{5}{|c|}{Without Payload}\\\hline
        0.001s  & 80.83 & 0.0012 & 0.001 & 0.048 \\\hline
        0.002s  & 46.18 & 0.0022 & 0.002 & 0.06 \\\hline
        \multicolumn{5}{|c|}{With Payload of 1Kg}\\\hline
        0.001s  & 75.88 & 0.0013 & 0.001 & 0.064 \\\hline
        0.002s  & 45.74 & 0.0022 & 0.002 & 0.11 \\\hline
    \end{tabular}
    \label{tab:petc_monitoring_period}
\end{table}
\begin{table}[!ht]
    \centering
    \caption{Steady-state RMSE $(t > T = 4 s)$ of the joint position (deg) and velocity (deg/s) for the FR3 robot with varying monitoring period  $h$, and with/without 1 kg payload. }
    \begin{tabular}{|c|c|c|c|c|c|c|c|}
    \hline
    \multicolumn{8}{|c|}{Without Payload}\\\hline
        $h$ (s) & $q_1$  & $q_2$  & $q_3$  & $q_4$  & $q_5$  & $q_6$  & $q_7$  \\ \hline
       0.001 & 0.088 & 0.062 & 0.099 & 0.080 & 0.098 & 0.105 & 0.100\\\hline
       0.002 & 0.095 & 0.066 & 0.103 & 0.084 & 0.098 & 0.106 & 0.097\\\hline
      $h$ (s) & $\dot{q}_1$  & $\dot{q}_2$  & $\dot{q}_3$  & $\dot{q}_4$  & $\dot{q}_5$  & $\dot{q}_6$ &  $\dot{q}_7$ \\\hline
       0.001 & 4.307 & 2.171 & 2.232 & 2.876 & 3.793 & 1.643 & 0.994\\\hline
       0.002 & 4.844 & 2.543 & 2.577 & 3.193 & 4.137 & 1.853 & 1.251\\\hline
       \multicolumn{8}{|c|}{With Payload of 1Kg}\\\hline
       $h$ (s) & $q_1$  & $q_2$  & $q_3$  & $q_4$  & $q_5$  & $q_6$  & $q_7$  \\ \hline
       0.001 & 0.088 & 0.114 & 0.106 & 0.112 & 0.130 & 0.087 & 0.099\\\hline
       0.002 & 0.094 & 0.150 & 0.109 & 0.115 & 0.120 & 0.084 & 0.097\\\hline
      $h$ (s) & $\dot{q}_1$  & $\dot{q}_2$  & $\dot{q}_3$  & $\dot{q}_4$  & $\dot{q}_5$  & $\dot{q}_6$ &  $\dot{q}_7$ \\\hline
       0.001 & 4.468 & 2.220 & 2.546 & 3.129 & 4.052 & 1.661 & 1.269\\\hline
       0.002 & 4.854 & 2.808 & 2.887 & 3.360 & 4.534 & 1.883 & 2.111\\\hline
    \end{tabular}
  
    \label{tab:petc_trackingError}
\end{table}
\begin{figure*}[!t] 
    \centering
    \subfloat[Joint Position $(\q)$]{\includegraphics[width=0.325\textwidth]{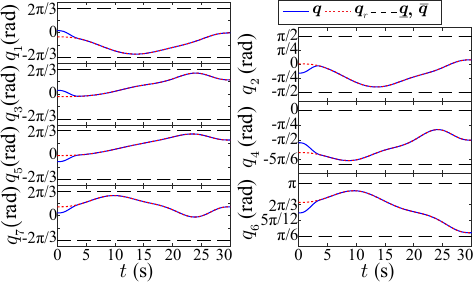}%
    \label{fig:petc_franka_jointPosition}}
    \hfill
    \subfloat[Joint Velocity $(\qd)$]{\includegraphics[width=0.325\textwidth]{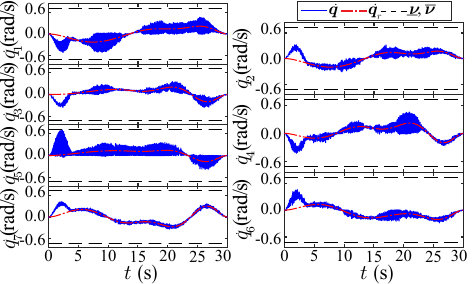}
    \label{fig:petc_franka_jointVelocity}}
    \hfill
    \subfloat[Control Input $(\bs{u})$]{\includegraphics[width=0.325\textwidth]{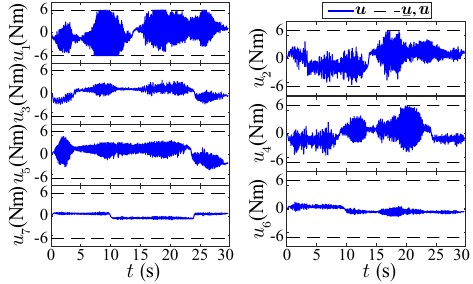}
    \label{fig:petc_franka_jointTorque}}
    \caption{Experimental results for the FR 3 robotic arm tracking a reference trajectory illustrating the satisfaction of state and input constraints using the proposed PETC scheme \policy{} with monitoring period $h = 0.001$s and prescribed time $T=4s$.}
\end{figure*}

\begin{figure*}[!t] 
    \centering
    \subfloat{\includegraphics[width=0.325\textwidth]{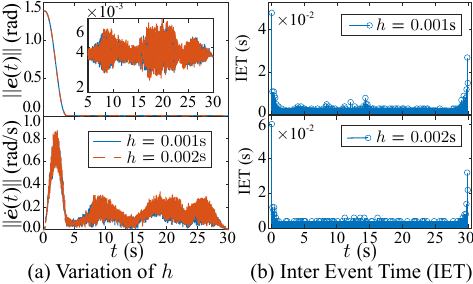}%
    \label{fig:petc_franka_variation_monitoring_period}}
    \hfill
    \subfloat{\includegraphics[width=0.325\textwidth]{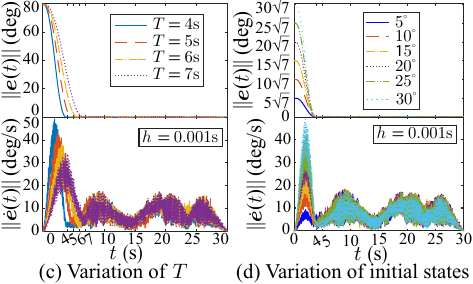}
    \hfill
    \subfloat{\includegraphics[width=0.325\textwidth]{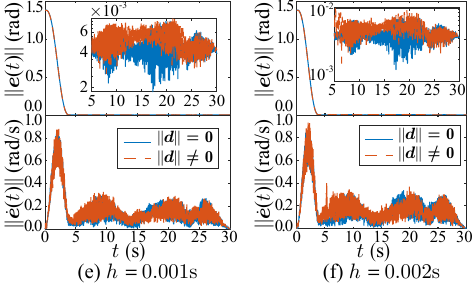}
    \label{fig:petc_franka_ext_disturbances}}
    \label{fig:petc_franka_variation_settling_time_inital_conditions}}
    \caption{Experimental results for the FR 3 robotic arm tracking a reference trajectory illustrating the satisfaction of state and input constraints using the proposed control policy \policy{}.}
    \label{fig:petc_franka_full_variation}
\end{figure*}
The effectiveness of the proposed PETC policy \policy{} is validated in this subsection through experimental evaluation on the Franka Research 3 (FR3), a 7-DoF serial robotic manipulator arm. The control loop rate for the FR3 robot is $1$KHz. For this experiment, the controller gains are chosen as: $\K {=} \text{diag}\paranthesis{[340, 324, 128, 224, 196, 64, 64]}$, $\rho {=} 4.5,\, \omega {=} 5.0,\ \gamma_1 {=} 1,\ \gamma_2 {=} 0.4,\ T{ =} 4\text{s}, \alpha {=} 0.03, \beta_0{=}0.0241$ and the state and input constraints are $\ol{\bs{q}} = [2.094, 1.5708, 2.094, 0.0, 2.094, 3.14, 2.094]^\top\text{rad}$, $\ul{\bs{q}} = [-2.094, -1.5708, -2.094, -2.86, -2.094, 0.5236, -2.094]^\top$, $\ol{\bs{\nu}} {=} - \ul{\bs{\nu}} {=} 0.6\!{\cdot}\!\bs{1}_7\ \text{rad/s}$ {and} $ \ol{\bs{{u}}} {=} -\ul{\bs{u}} =  6\!{\cdot}\!\bs{1}_7\ \text{Nm}$, respectively.

Firstly, we consider a minimum jerk trajectory \cite{adam:2015:trajGen} as a reference, and an initial offset of $30^\circ$ in each joint from the reference is considered as the initial state together with $\qd(0) = \bs{0}_7$. The monitoring period is chosen as $h = 0.001$s. Fig. \ref{fig:petc_franka_jointPosition}, and \ref{fig:petc_franka_jointVelocity} demonstrate that the control policy \policy{} does not violate state constraints even while tracking the desired trajectory. Furthermore, with the nonlinear transformation of the input signal that invokes the use of the saturation function \eqref{eq:petc_saturation_function}, the control policy \policy{} satisfies the input constraints as shown in Fig. \ref{fig:petc_franka_jointTorque}. Moreover, the steady-state performance shown in Table \ref{tab:petc_trackingError} illustrates that the RMSE in joint position and velocity is on the order of $0.1$ deg and $5$ deg/s, respectively, when no external payload is attached. Consequently, the steady-state metrics showcase the efficacy of the proposed policy after the prescribed time convergence has been achieved.

In addition, an experimental verification for different values of the monitoring period is carried out as shown in Fig. \ref{fig:petc_franka_full_variation}a and \ref{fig:petc_franka_full_variation}b. Note that with the PETC mechanism \eqref{eq:petc_mechanism}, there is no Zeno behaviour exist, resulting in MIET $0.001$s and $0.002$s with corresponding monitoring period $h=0.001$s and $h=0.002$s, respectively. As evidenced in Table \ref{tab:petc_monitoring_period}, the transmission frequency of the control updates when deploying the PETC mechanism \eqref{eq:petc_mechanism} drops to about 80\% for the case of $h = 0.001$s. On the other hand, with a monitoring period of $h=0.002s$, about 46\% of control updates are only required to achieve prescribed time tracking within the prescribed bound. Furthermore, the performance of the robot with a 1 kg payload is also undertaken while varying the monitoring period, as shown in Figs. \ref{fig:petc_franka_full_variation}e and \ref{fig:petc_franka_full_variation}f. 

 Lastly, tracking performance for different initial states and prescribed settling-times is also shown in Fig. \ref{fig:petc_franka_full_variation}c and \ref{fig:petc_franka_full_variation}d. Clearly, for a range of initial conditions and user-defined settling times, the control policy \policy{} achieves local PTPB convergence while strictly adhering to state and input constraints. Consequently, the proposed control policy \policy{} along with the PETC mechanism \eqref{eq:petc_mechanism} demonstrates that it can reject disturbances robustly while tracking a reference trajectory with reduced control updates.

\section{Conclusion} \label{sec:conclusion}
This article proposes a periodic event-triggered adaptive barrier control policy for addressing the trajectory tracking problem of EL systems with SIT constraints. In particular, an approximation-free adaptive-barrier control architecture is designed to ensure the convergence of tracking error within a prescribed time frame. Furthermore, the PETC strategy alleviates the need for continuous monitoring of the event-triggering mechanism, thereby reducing the communication bandwidth required for control signals. Moreover, the stability analysis formally verifies that the proposed controller guarantees convergence of all tracking errors to the prescribed bounds under SIT constraints. Also, this study provides a detailed analysis of the PETC mechanism, showing that the choice of monitoring period affects the tracking performance. Finally, numerical and experimental results are presented to demonstrate the superior performance of the proposed scheme. As part of future work, we plan to extend this study to trajectory tracking for fully actuated and under-actuated EL systems with vision in the loop.


\begin{thebibliography}{10}
	\providecommand{\url}[1]{#1}
	\csname url@samestyle\endcsname
	\providecommand{\newblock}{\relax}
	\providecommand{\bibinfo}[2]{#2}
	\providecommand{\BIBentrySTDinterwordspacing}{\spaceskip=0pt\relax}
	\providecommand{\BIBentryALTinterwordstretchfactor}{4}
	\providecommand{\BIBentryALTinterwordspacing}{\spaceskip=\fontdimen2\font plus
		\BIBentryALTinterwordstretchfactor\fontdimen3\font minus \fontdimen4\font\relax}
	\providecommand{\BIBforeignlanguage}[2]{{%
			\expandafter\ifx\csname l@#1\endcsname\relax
			\typeout{** WARNING: IEEEtran.bst: No hyphenation pattern has been}%
			\typeout{** loaded for the language `#1'. Using the pattern for}%
			\typeout{** the default language instead.}%
			\else
			\language=\csname l@#1\endcsname
			\fi
			#2}}
	\providecommand{\BIBdecl}{\relax}
	\BIBdecl
	
	\bibitem{obuz2024robust}
	S.~Obuz, E.~Selim, E.~Tatlicioglu, and E.~Zergeroglu, ``Robust prescribed time control of euler--lagrange systems,'' \emph{IEEE Transactions on Industrial Electronics}, 2024.
	
	\bibitem{he2023ude}
	S.~He, S.-L. Dai, Z.~Zhao, T.~Zou, and Y.~Ma, ``Ude-based distributed formation control for msvs with collision avoidance and connectivity preservation,'' \emph{IEEE Transactions on Industrial Informatics}, vol.~20, no.~2, pp. 1476--1487, 2023.
	
	\bibitem{lu2019adaptive}
	M.~Lu, L.~Liu, and G.~Feng, ``Adaptive tracking control of uncertain euler--lagrange systems subject to external disturbances,'' \emph{Automatica}, vol. 104, pp. 207--219, 2019.
	
	\bibitem{he2023output}
	X.~He and M.~Lu, ``Output feedback control of uncertain euler--lagrange systems by internal model,'' \emph{Automatica}, vol. 156, p. 111189, 2023.
	
	\bibitem{feng2023event}
	Y.~Feng, L.~Kong, Z.~Zhang, R.~Liu, G.~Cheng, and X.~Yu, ``Event-triggered finite-time control for a constrained robotic manipulator with flexible joints,'' \emph{International Journal of Robust and Nonlinear Control}, vol.~33, no.~11, pp. 6031--6051, 2023.
	
	\bibitem{chen2025fixed}
	Z.~Chen, Z.~Li, Z.~Xiong, W.~Han, and J.~Wang, ``Fixed-time periodic adaptive event-triggered control for robotic manipulator,'' \emph{International Journal of Adaptive Control and Signal Processing}, 2025.
	
	\bibitem{polykao_FT_EXPAP}
	F.~Lopez-Ramirez, D.~Efimov, A.~Polyakov, and W.~Perruquetti, ``Finite-time and fixed-time input-to-state stability: Explicit and implicit approaches,'' \emph{Systems \& Control Letters}, vol. 144, p. 104775, 2020.
	
	\bibitem{song2023prescribed}
	Y.~Song, H.~Ye, and F.~L. Lewis, ``Prescribed-time control and its latest developments,'' \emph{IEEE Transactions on Systems, Man, and Cybernetics: Systems}, vol.~53, no.~7, pp. 4102--4116, 2023.
	
	\bibitem{zhang2023sampled}
	X.-M. Zhang, Q.-L. Han, X.~Ge, B.~Ning, and B.-L. Zhang, ``Sampled-data control systems with non-uniform sampling: A survey of methods and trends,'' \emph{Annual Reviews in Control}, vol.~55, pp. 70--91, 2023.
	
	\bibitem{zhang2023recent}
	P.~Zhang, T.~Liu, J.~Chen, and Z.-P. Jiang, ``Recent developments in event-triggered control of nonlinear systems: An overview,'' \emph{Unmanned Systems}, vol.~11, no.~01, pp. 27--56, 2023.
	
	\bibitem{borgers2014event}
	D.~P. Borgers and W.~M.~H. Heemels, ``Event-separation properties of event-triggered control systems,'' \emph{IEEE Transactions on Automatic Control}, vol.~59, no.~10, pp. 2644--2656, 2014.
	
	\bibitem{benitez2020periodic}
	S.~E. Benitez-Garcia, M.~G. Villarreal-Cervantes, J.~F. Guerrero-Castellanos, and J.~P. S{\'a}nchez-Santana, ``Periodic event-triggered control for the stabilization of robotic manipulators,'' \emph{IEEE Access}, vol.~8, pp. 111\,553--111\,565, 2020.
	
	\bibitem{eqtami2010event}
	A.~Eqtami, D.~V. Dimarogonas, and K.~J. Kyriakopoulos, ``Event-triggered control for discrete-time systems,'' in \emph{Proceedings of the 2010 american control conference}.\hskip 1em plus 0.5em minus 0.4em\relax IEEE, 2010, pp. 4719--4724.
	
	\bibitem{gao2022novel}
	J.~Gao, ``A novel event-triggered adaptive tracking control framework for a manipulator with aperiodic neural network estimation,'' \emph{Assembly Automation}, vol.~42, no.~4, pp. 411--426, 2022.
	
	\bibitem{dang2024event}
	B.~Dang and H.~Li, ``Event-triggered stochastic finite-time tracking control of robot manipulator with uncertain disturbance neural network estimation,'' \emph{Nonlinear Dynamics}, vol. 112, no.~18, pp. 16\,315--16\,337, 2024.
	
	\bibitem{kumari2018event}
	K.~Kumari, A.~K. Behera, and B.~Bandyopadhyay, ``Event-triggered sliding mode-based tracking control for uncertain euler--lagrange systems,'' \emph{IET Control Theory \& Applications}, vol.~12, no.~9, pp. 1228--1235, 2018.
	
	\bibitem{chen2024robust}
	L.~Chen and F.~Hao, ``Robust tracking control for uncertain euler--lagrange systems via dynamic event-triggered and self-triggered adp,'' \emph{International Journal of Robust and Nonlinear Control}, vol.~34, no.~1, pp. 481--505, 2024.
	
	\bibitem{chen2024event}
	C.~Chen, Z.~Peng, C.~Zou, R.~Huang, K.~Shi, and H.~Cheng, ``Event-triggered learning-based robust tracking control for robotic manipulators with uncertain dynamics and non-zero equilibrium,'' \emph{Expert Systems with Applications}, vol. 255, p. 124573, 2024.
	
	\bibitem{peng2023event}
	Z.~Peng, W.~Yan, R.~Huang, H.~Cheng, K.~Shi, and B.~K. Ghosh, ``Event-triggered learning robust tracking control of robotic systems with unknown uncertainties,'' \emph{IEEE Transactions on Circuits and Systems II: Express Briefs}, vol.~70, no.~7, pp. 2540--2544, 2023.
	
	\bibitem{gao2022observer}
	J.~Gao, W.~He, and H.~Qiao, ``Observer-based event and self-triggered adaptive output feedback control of robotic manipulators,'' \emph{International Journal of Robust and Nonlinear Control}, vol.~32, no.~16, pp. 8842--8873, 2022.
	
	\bibitem{soni2022sliding}
	S.~K. Soni, S.~Kumar, S.~Wang, A.~Sachan, D.~Boutat, and D.~Geha, ``Sliding mode even-triggered tracking control for robot manipulators with state constrains,'' in \emph{IECON 2022--48th Annual Conference of the IEEE Industrial Electronics Society}.\hskip 1em plus 0.5em minus 0.4em\relax IEEE, 2022, pp. 1--6.
	
	\bibitem{zhang2023event}
	Z.~Zhang, Y.~Gao, W.~Sun, and Y.~Wu, ``Event-based nonsingular fixed-time tracking control of an uncertain manipulator system subject to full-state static constraints,'' \emph{IEEE Transactions on Cybernetics}, 2023.
	
	\bibitem{feng2024event}
	X.~Feng and X.~Zhang, ``Event-triggered shared control for euler--lagrange systems with actuator fault and unknown control direction,'' \emph{Asian Journal of Control}, 2024.
	
	\bibitem{li2022finite}
	C.~Li, L.~Zhao, and Z.~Xu, ``Finite-time adaptive event-triggered control for robot manipulators with output constraints,'' \emph{IEEE Transactions on Circuits and Systems II: Express Briefs}, vol.~69, no.~9, pp. 3824--3828, 2022.
	
	\bibitem{liu2021event}
	W.~Liu and J.~Huang, ``Event-triggered exponential practical tracking for uncertain euler-lagrange systems,'' in \emph{2021 American Control Conference (ACC)}.\hskip 1em plus 0.5em minus 0.4em\relax IEEE, 2021, pp. 5003--5008.
	
	\bibitem{sui:2025:FTC}
	J.~Sui, B.~Niu, Y.~Ou, X.~Zhao, and D.~Wang, ``Event-triggered adaptive finite-time control for a robotic manipulator system with global prescribed performance and asymptotic tracking,'' \emph{IEEE Transactions on Cybernetics}, vol.~55, no.~3, pp. 1045--1055, 2025.
	
	\bibitem{liu2024adaptive}
	D.~Liu, X.~Ouyang, N.~Zhao, and Y.~Luo, ``Adaptive event-triggered control with prescribed performance for nonlinear system with full-state constraints.'' \emph{IAENG International Journal of Applied Mathematics}, vol.~54, no.~3, 2024.
	
	\bibitem{ning:2023:scalar_input}
	P.~Ning, C.~Hua, K.~Li, and R.~Meng, ``Event-triggered control for nonlinear uncertain systems via a prescribed-time approach,'' \emph{IEEE Transactions on Automatic Control}, vol.~68, no.~11, pp. 6975--6981, 2023.
	
	\bibitem{yang:2024:scalar_input}
	H.~Yang, Y.~Wang, and Z.~Shao, ``Event-triggered prescribed-time control for a class of uncertain nonlinear systems using finite time-varying gain,'' \emph{ISA Transactions}, vol. 152, pp. 167--176, 2024.
	
	\bibitem{jiang2024event}
	T.~Jiang, Y.~Yan, S.~Yu, T.~Li, and H.~Sang, ``Event-triggered based predefined-time tracking control for robotic manipulators with state and input quantization,'' \emph{Nonlinear Dynamics}, vol. 112, no.~21, pp. 19\,169--19\,183, 2024.
	
	\bibitem{chen2023adaptive}
	Z.~Chen, H.~Zhang, J.~Liu, Q.~Wang, and J.~Wang, ``Adaptive prescribed settling time periodic event-triggered control for uncertain robotic manipulators with state constraints,'' \emph{Neural Networks}, vol. 166, pp. 1--10, 2023.
	
	\bibitem{Dingxin:2021:icra}
	D.~He, H.~Wang, Y.~Tian, and Y.~Tao, ``Event-triggered model-free prescribed-time quantized control for second-order nonlinear system with uncertainties and external disturbances,'' in \emph{2021 6th International Conference on Robotics and Automation Engineering (ICRAE)}, 2021, pp. 353--357.
	
	\bibitem{wang2022adaptive}
	Z.~Wang, H.-K. Lam, Y.~Guo, B.~Xiao, Y.~Li, X.~Su, E.~M. Yeatman, and E.~Burdet, ``Adaptive event-triggered control for nonlinear systems with asymmetric state constraints: A prescribed-time approach,'' \emph{IEEE Transactions on Automatic Control}, vol.~68, no.~6, pp. 3625--3632, 2022.
	
	\bibitem{hu2022event}
	Y.~Hu, H.~Yan, Y.~Zhang, H.~Zhang, and Y.~Chang, ``Event-triggered prescribed performance fuzzy fault-tolerant control for unknown euler--lagrange systems with any bounded initial values,'' \emph{IEEE Transactions on Fuzzy Systems}, vol.~31, no.~6, pp. 2065--2075, 2022.
	
	\bibitem{csk:2025}
	C.~S. Kashyap, P.~Jagtap, and J.~Keshavan, ``Tracking control of euler-lagrangian systems with prescribed state, input, and temporal constraints,'' \emph{arXiv preprint arXiv:2503.01866}, 2025.
	
	\bibitem{kgarg_clf}
	K.~Garg, E.~Arabi, and D.~Panagou, ``Prescribed-time convergence with input constraints: A control lyapunov function based approach,'' in \emph{2020 American Control Conference (ACC)}, 2020, pp. 962--967.
	
	\bibitem{PTC_TBG}
	G.~Arechavaleta, J.~Obreg{\'o}n, H.~M. Becerra, and A.~Morales-D{\'\i}az, ``Predefined-time convergence in task-based inverse dynamics using time base generators,'' \emph{IFAC-PapersOnLine}, vol.~51, no.~13, pp. 443--449, 2018.
	
	\bibitem{zhang2008repetitive}
	Y.~Zhang, X.~Lv, Z.~Li, Z.~Yang, and K.~Chen, ``Repetitive motion planning of pa10 robot arm subject to joint physical limits and using lvi-based primal--dual neural network,'' \emph{Mechatronics}, vol.~18, no.~9, pp. 475--485, 2008.
	
	\bibitem{adam:2015:trajGen}
	A.~Bry, C.~Richter, A.~Bachrach, and N.~Roy, ``Aggressive flight of fixed-wing and quadrotor aircraft in dense indoor environments,'' \emph{The International Journal of Robotics Research}, vol.~34, no.~7, pp. 969--1002, 2015.
	
\end{thebibliography}
\end{document}